\documentclass[12pt,aps,preprint,pra,amsmath,amssymb]{revtex4}
\usepackage{epsfig}

\def\la{\langle}
\def\ra{\rangle}

\def\da{^\dagger}

\newcommand{\Op}[1]{{\boldsymbol{\mathrm{\widehat{#1}}}}}

\begin{document}

\title{Optimal control theory for unitary transformations}

\author{Jos\'e P. Palao $^{(a,b)}$ and Ronnie Kosloff $^{(a)}$}

\affiliation{$^{(a)}$ Department of Physical Chemistry and the Fritz 
Haber Research Center for Molecular Dynamics, Hebrew University, 
Jerusalem 91904, Israel\\
$^{(b)}$ Departamento de F\'{\i}sica Fundamental II, Universidad de 
La Laguna, La Laguna 38204, Spain}


\begin{abstract}
The dynamics of a quantum system driven by an external field is well
described by a unitary transformation generated by a time dependent Hamiltonian.
The inverse problem of finding the field that generates a specific unitary
transformation is the subject of study.
The unitary transformation which 
can represent an algorithm in a quantum computation is imposed on
a subset of quantum states embedded in a larger Hilbert space. 
Optimal control theory (OCT) is used to solve the inversion problem
irrespective of the initial input state. A unified formalism, based on 
the Krotov method is developed leading to a new scheme. 
The schemes are compared for the inversion of a two-qubit Fourier transform
using as registers the vibrational levels of the $X^1\Sigma^+_g$ 
electronic state of Na$_2$. 
Raman-like  transitions through the $A^1\Sigma^+_u$  electronic state 
induce the transitions. Light fields are found that are 
able to implement the Fourier transform within a picosecond time scale.
Such fields can be obtained by pulse-shaping techniques of a femtosecond pulse.
Out of the schemes studied the square modulus scheme converges fastest.
A study of the implementation of the $Q$ qubit
Fourier transform in the Na$_2$ molecule was carried out for up to 5 qubits. 
The classical computation effort required to obtain the algorithm
with a given fidelity is estimated to scale exponentially with the number of 
levels. The observed moderate scaling of the pulse intensity with the number of qubits
in the transformation is rationalized.

\noindent
\\
\noindent{PACS number(s): 82.53.Kp 03.67.Lx
33.90.+h 32.80.Qk}
\\
\end{abstract}

\maketitle

\section{Introduction}\label{se:introduction}

Coherent control was initiated to steer 
a quantum system to a final objective via an external field \cite{RZ00,SB03}.
If the initial and final objective states are pure, the method 
can be termed state-to-state coherent control. 
By  generalizing, the problem  of
steering simultaneously a {\em set} of initial pure states 
to a set of final states can be formulated.
Such a possibility  has direct applicability in quantum computing
where an algorithm implemented as a unitary transformation operating 
on a set of states  has to be carried out irrespective of the input.
In this application both input and output are 
encoded as a superposition of these states \cite{NC00}.
  
To implement such a control, the external driving field that 
induces a pre-specified unitary transformation has to be found. 
Different methods have been suggested for this task. Some rely on 
factorizing the algorithm encoded as a unitary transformation, to a set
of elementary gates and then finding a control solution for the
elementary unitary evolution of a single gate \cite{NC00,lloyd95}. 
The inherent difficulty in such an approach is that in general  the field 
addresses many levels simultaneously. Therefore, when  
a particular single gate operation is carried out other levels are affected.
This means that the ideal single gate unitary transformation has to be implemented 
so that all other possible transitions are avoided.
The problem is simpler when each allowed transition
is selectively addressable \cite{SGRR02}.  However, in general
the problem of undesired coupling has to be corrected.
A specific solution has been suggested  \cite{TL00} 
but a general solution is not known. 

The presence of a large number of levels coupled to the 
external driving field is especially relevant in the
implementation of quantum computing in molecular systems
\cite{ZKLA01,AKL02,TKV01}.
The use of optimal control theory (OCT) has been proposed 
as a possible solution \cite{TKV01,RB01}. 
In recent years OCT for quantum systems \cite{RZ00}  has received considerable
attention, leading to effective methods for obtaining 
the driving field which will induce a desired transition 
between preselected initial and final states. 
To address the control problem of inducing a particular unitary
transformation the state-to-state OCT has to be augmented.
For example, if the unitary transformation is to relate the initial states 
$|\varphi_{ik}\rangle$ with the final states 
$|\varphi_{fk}\rangle$,
the state-to-state OCT  derives an optimal field $\epsilon_k$
for each pair of initial and final states $(\varphi_{ik},\varphi_{fk})$.
But the fields $\epsilon_k$ obtained are in general different, 
so that the evolution induced by $\epsilon_k$ is not
appropriate for a different set of initial and final states.
In order to implement a given unitary transformation a single field
$\epsilon$ that relates simultaneously to all the relevant pairs 
$(\varphi_{ik},\varphi_{fk})$ is needed.

Two approaches  have been suggested to generalize OCT for 
unitary transformations. The first approach 
is formulated directly on the evolution operator \cite{PK02}. 
An alternative approach uses the 
simultaneous optimization of several state-to-state transitions
\cite{RB01,TV02}. 
The present paper develops a comprehensive framework 
for constructing an OCT solution for the unitary transformation.
The study explores various approaches. A common framework for
an iterative solution based on the Krotov approach \cite{TKO92} is developed. 
As a result, the numerical implementation
of the methods  are almost identical enabling an unbiased assessment.
The implementation of the Fourier transform algorithm in a molecular
environment is chosen as a case study. 
The performance of the various OCT schemes is compared in a realistic setup.
A crucial demand in quantum computing is obtaining  
high fidelity of the solution. The present OCT scheme can be viewed as
an iterative  classical algorithm which finds a field that implements
the quantum algorithm. The obvious questions are:
\begin{itemize}
\item{What are the
computational resources required to obtain a high fidelity result?}
\item{How do these computational resources scale with the number of qubits in the
quantum algorithm?}
\item{How do the actual physical resources i.e. the integrated power of the field
scale with the number of qubits in the quantum algorithm?}
\end{itemize}

The paper is organized as follow: In Sec. \ref{se:implementation} 
the problem is formulated introducing different 
objectives devoted to the optimization of a given 
unitary transformation. 
In Sec. \ref{se:optimization} the application of the Krotov method of 
optimization of the given objectives is described. Expressions obtained for the
optimal field are formulated as well as the implementation of the method. 
The variational method to derive the optimization 
equations is commented on in Appendix \ref{ap:varapp}.
The results are used to study the implementation of  a unitary transformation in 
a molecular model  Sec. \ref{se:qubits}.
Finally, in Sec. \ref{se:conclusions}
results are discussed.


\section{Implementation of a unitary transformation}
\label{se:implementation}

\subsection{Description of the problem}

The objective of the study is to devise a method to find the driving field
that executes a unitary transformation on a subsystem embedded in a larger Hilbert space.
The unitary transformation is to be applied in a Hilbert 
space $\cal M$ of dimension $M$, expanded by an orthonormal 
basis of states $\{|m\ra\}$ ($m=1,...,M$). 
The selected unitary transformation is imposed on the subspace
$\cal N$ of $N$ levels of the system ($N\leq M$).
In the context of quantum computation, the $N$ 
levels correspond to the physical implementation of the 
qubit(s) embedded in a larger system.
The additional levels ($m=N+1,...,M$), considered as ``spurious levels'',
are only indirectly involved in the target unitary transformation. 

In any realistic implementation of quantum computing,
"spurious levels" always exist. One reason is that the system
is never completely isolated from the environment.
In addition, the control levers, that in the present case is 
the dipole operator, connects  directly only part of the primary levels.
An example is the implementation of quantum
computation using rovibronic molecular levels.
The transition dipole connects two electronic surfaces \cite{PK02}. 
The primary states reside on one surface, so that Raman-like transitions
are used to implement the unitary transformation.
The advantage of this setup is that the
transitions frequencies between the electronic surfaces are in the visible 
region, for which the pulse shaping technology is well
developed \cite{nelson}. Other 
levels residing on both of the electronic surfaces become spurious in the sense
that any leakage to them destroys the desired final
result. However at intermediate times these levels constitute a temporary 
storage space which facilitates the execution of the transformation.

The objective is to implement a selected 
unitary transformation in the relevant
subspace ${\cal N}$ at a given final time $T$.
The target unitary transformation is represented by an
operator in the Hilbert space of the primary system
and is denoted by $\Op{O}$. 
For $N<M$, the matrix  representation of $\Op{O}$ 
in the basis $\{|m\ra\}$ has two blocks of dimension 
$N\times N$ and $(M-N)\times(M-N)$. The elements
connecting these blocks are zero. 
This structure means that population  
is not transferred between the two
subspaces at the target time $T$,
but can take place at intermediate times. 
Only the $N\times N$ block is relevant for the 
optimization procedure, while the other remains arbitrary.

The dynamics of the system is generated by the Hamiltonian
$\Op{H}$,
\begin{equation}\label{eq:Hamiltonian}
\Op{H}(\epsilon)=\Op{H}_0-\Op{\mu}\,\epsilon(t)\,,
\end{equation}
where $\Op{H}_0$ is the free Hamiltonian, $\epsilon(t)$ is
the driving field and $\Op{\mu}$ is a system operator describing 
the coupling between system and field. In the molecular systems,
this coupling corresponds to the transition dipole operator and the 
driving field becomes radiation. 
In some cases more than one independent driving field can 
be considered. An example is when two 
components of the polarization of an electro-magnetic 
field are separately controlled \cite{BG01}. The generalization of the
formalism in such a case is straightforward.

The system dynamics is fully specified 
by the evolution operator $\Op{U}(t,0;\epsilon)$.
An optimal field $\epsilon_{opt}$ induces the
target unitary transformation $\Op{O}$ at time $T$ 
when
\begin{equation}\label{eq:condition}
\Op{U}(T,0;\epsilon_{opt})\,=\,e^{-i\phi(T)}\,\Op{O}\,.
\end{equation}
Eq. (\ref{eq:condition}) implies a condition on only the 
$N\times N$ block of the matrix representation of $\Op{U}$.
The phase $\phi(T)$ is introduced to point out that 
the target unitary transformation $\Op{O}$ can
be implemented only up to an arbitrary global phase.
The phase $\phi$ can be decomposed into two terms, $\phi_1(T)+\phi_2(T)$.
The first, $\phi_1$, originates from the arbitrary choice of the
origin of the energy levels. Formally, a term
proportional to the identity operator can always be added to the
Hamiltonian. 
When the states $|m\rangle$ correspond to the eigenstates of
$\Op{H}_0$, the phase $\phi_1$ is
\begin{equation}\label{eq:phase}
\phi_{1}(T)=\frac{\sum_{m=1}^M\,E_m\,T}{M\,\hbar}\,,
\end{equation}
where $E_m$ is the energy of the level $m$.
The phase $\phi_2$ reflects the arbitrariness 
of the unitary transformation for the levels $m=N+1,...,M$
which are not part of the target.

The method to determine the optimal field is based on maximizing
a real functional of the field that
depends on both the target unitary transformation and the 
evolution generated by the Hamiltonian, fulfilling Eq. 
(\ref{eq:condition}). 
The problem of unitary transformation optimization is then 
reduced to a functional optimization.
However different formulations of the problem can be made, 
leading to different functionals and then, 
in principle to different results.
In the present context two different formulations have been proposed,
one based on the evolution operator 
\cite{PK02} and the other on simultaneous $N$ 
state to state transitions \cite{TV02}. 
These formulations are closely related. A similar optimization 
procedure  is described in Sec. \ref{se:optimization}.

\subsection{Evolution operator formulation}

The optimization formulation is based on the definition of 
a complex functional $\tau$ that depends on the 
evolution operator at time $T$ \cite{PK02}.
The following functional is introduced:
\begin{equation}\label{eq:tau}
\tau(\Op{O};T;\epsilon)
={\rm Tr}\{\Op{O}\da\Op{U}(T,0;\epsilon)\Op{P}_N\}\,
=\sum_{n=1}^N \la n|\Op{O}\da\Op{U}(T,0;\epsilon)|n\ra\,,
\end{equation}
where the projection $\Op{P}_N=\sum_{n=1}^N\,|n\ra\la n|$ is used.
$\{|n\ra\}$ denotes an orthonormal basis
of the subspace $\cal N$. 
As $\Op{O}$ is a unitary transformation in the relevant
subspace, the functional $\tau$ is a complex number restricted to the interior
of a circle in the complex plane of radius $N$ centered at the origin. 
The modulus of $\tau$ is equal
to $N$ only for an optimal field fulfilling Eq. (\ref{eq:condition}).
$\tau$ can then be interpreted as an indicator  of the
fidelity of the implementation on the target unitary transformation  
\cite{PK02}. When $\tau $ approaches  $N$,
the transformation imposed by the field converges to
the target objective.

Since $\tau$ is complex, several different real functionals 
can be associated with it.
In Ref. \cite{PK02} the optimization of the real part 
of $\tau$, or the imaginary part, or a linear combination 
of both was suggested to find the optimal field.
It was found that all these possibilities show a similar 
performance. For this reason, the present paper employs the optimization 
of the real part chosen as a representative case.  
The functional is therefore defined as:
\begin{equation}\label{eq:Fre}
F_{re}=
-{\rm Re}\left[\tau(\Op{O};T;\epsilon)\right]=
-{\rm Re}\left[
\sum_{n=1}^{N}\la n|\Op{O}\da\Op{U}(T,0;\epsilon)|n\ra\right]\,.
\end{equation}
The functional reaches its minimum value, $F_{re}=-N$, when the 
driving field induces the target unitary transformation but however
with the additional condition that the phase term $\exp(-i\phi(T))$ is equal 
to one. 

Other functionals based on $\tau$ but without any
condition on the phase can be defined. In this work
the squared modulus of $\tau$ with a negative sign is studied:
\begin{equation}\label{eq:Fsm}
F_{sm}=
-|\tau(\Op{O};T;\epsilon)|^2=
-\sum_{n=1}^{N}\sum_{n^\prime=1}^{N}
\la n|\Op{O}\da\Op{U}(T,0;\epsilon)|n\ra\,
\la n^\prime|\Op{U}(T,0;\epsilon)\da\Op{O}|n^\prime\ra\,,
\end{equation}
with minimum value $F_{sm}=-N^2$ for any field satisfying 
Eq. (\ref{eq:condition}).

\subsection{Formulation of the Simultaneous $N$ state to state transitions}

This formulation is based on the
simultaneous optimization of $N$ transitions between a set
of initial states $|l\rangle$ and the corresponding
final states $\Op{O}|l\rangle$ ($l=1,\dots,N$) \cite{TV02}. 
For this
purpose the following functional is defined,
\begin{equation}\label{eq:eta}
\eta(\Op{O};T;\epsilon)
={\rm Tr}\left\{\sum_{l=1}^{N}\Op{O}\da\Op{U}(T,0;\epsilon)
| l\ra\,\la\l|
\Op{U}(T,0;\epsilon)\da\Op{O}
|l\ra\,\la\l|\right\} 
=\sum_{l=1}^{N} 
|\la l|\Op{O}\da\Op{U}(T,0;\epsilon)
| l\ra|^2\,.
\end{equation}
Notice that
while $\tau$ is defined as the sum of amplitudes, $\eta$ is defined
as the sum of overlaps at the final time $T$. The parameter $\eta$ is 
a positive real number and its  maximum value, $N$, 
is reached when all the initial states $|l\ra$
are driven by the field to the final target states $\Op{O}|l\ra$,
except for a possible arbitrary phase associated with each transition. 
The arbitrariness of these phases implies that
the set of initial states $|l\ra$ must be chosen
carefully. In order to account for all the possible transitions,
the states $|l\ra$ have to faithfully represent all the relevant 
subspace $\cal N$ i.e. constitute a complete basis set.
However, a choice of an orthonormal basis 
could produce undesired results. For example, the ambiguity
of using an orthonormal basis $\{|n\ra\}$
in the relevant subspace and  an arbitrary 
unitary transformation $\Op{D}$, diagonal in that basis. The product 
$\Op{O}\,\Op{D}$ will also be a unitary transformation. If 
$\epsilon_O$ and $\epsilon_{OD}$ are fields that
generate $\Op{O}$ and $\Op{O}\,\Op{D}$ at time $T$ respectively,
they both will have the same fidelity $\eta$,
\begin{equation}
\eta_\perp(\Op{O};T;\epsilon_O)=\eta_\perp(\Op{O};T;\epsilon_{OD})\,,
\end{equation}
where $\perp$ denotes that $\eta$ was evaluated using an orthonormal 
basis.
Then any algorithm based on optimizing $\eta$ that uses an orthonormal 
basis could find a solution for the field corresponding to 
the implementation of an arbitrary
unitary transformations of the form $\Op{O}\,\Op{D}$.
($\Op{O}$ is a particular case when $\Op{D}$ is the identity operator).
The reason for this discrepancy is that $\eta$ is only sensitive to the overlap of
each pair of initial $|l\ra$ and final $\Op{O}|l\ra$ states,
leaving undetermined the relative phases between them.
For the optimization procedure to succeed  a careful choice 
of the initial set of states is necessary. A simple solution is
to compose the  last $N$ state as a superposition of all states in the basis
$\sum_{l=1}^{N}|n\ra/\sqrt{N}$, and to 
keep as is the first $N-1$ states of an orthonormal basis.
For this set of states, the maximum condition is achieved only when the
field induces the target unitary transformation up to
a possible global phase. 

To summarize  the functional $-\eta$ in is used,
\begin{equation}\label{eq:Fss}
F_{ss}=-\eta(\Op{O};T;\epsilon),
\end{equation}
with a minimum value $F_{ss}=-N$. The optimal field reached 
satisfies Eq. (\ref{eq:condition}), 
subject to a choice of the set of states  $|l\ra$ which determines the
relative phases.

\subsection{Initial to final state optimization}

The present formulations of quantum control
assume that the target unitary transformation $\Op{O}$
is explicitly known, at least in the subspace $\cal N$.
In most previous applications of optimal control theory, the objective 
was specified as the maximization of the expectation 
value of a given observable at time $T$ subject to a predefined initial state 
\cite{RZ00}. 
Both mixed and pure initial states were considered \cite{BKT01,Ohtsuki03}.
A particular case is the determination of an optimal field 
to drive the system from a given pure initial state 
$|\varphi_i\rangle$ to a target pure final state 
$|\varphi_f\rangle$ at time $T$.
This state-to-state objective optimization
can be derived from the present formulation
if the target unitary transformation becomes
$\Op{O}|\varphi_i\rangle=|\varphi_f\rangle$. 
The evolution operator formulation is then obtained by setting the
projector 
$\Op{P}_N=|\varphi_i\rangle\langle\varphi_i|$ in Eq. (\ref{eq:tau}), 
obtaining the functional $\tau$,
\begin{equation}\label{eq:tau2}
\tau(\varphi_i,\varphi_f;T;\epsilon)
\,=\,\langle\varphi_f|\Op{U}(T,0;\epsilon)|\varphi_i\rangle\,.
\end{equation}
The real functionals $F_{re}=-{\rm Re}[\tau]$ 
and $F_{sm}=-|\tau|^2$ are to be used in the study.
As only a state-to-state transition is involved, the formulation
is obtained by choosing $|l\rangle\equiv|\varphi_i\rangle$.
In this case $\eta=|\tau|^2$ and $F_{ss}=F_{sm}$. Notice that
this result is valid only when there is a single term in the sum
in Eq. (\ref{eq:tau}) and in Eq. (\ref{eq:eta}).


\section{Optimization}\label{se:optimization}

A common optimization procedure for all the functionals
$F$ as defined in the previous section is developed.
The notation $|n\rangle$ for the states and $n$ its index will be used in the
evolution operator formulation. The notation $|l\rangle$ and $l$ will be used 
in the simultaneous $N$ state to state transitions formulation.  The notation
$|\varphi_{ik}\rangle$ and $k$ where $k=1,\dots,N$,
will be used when the results are valid for both cases.
An evaluation of any of the functionals requires the knowledge of the states 
$|\varphi_k(T)\rangle=\Op{U}(T,0;\epsilon)|\varphi_{ik}\rangle$
and $|\varphi_{fk}\rangle=\Op{O}|\varphi_{ik}\rangle$. 
The operation of the evolution equation  $\Op{U}(t,0;\epsilon)|\varphi_{ik}\rangle$
can be calculated by solving the time-dependent Schr\"odinger equation
\begin{equation}\label{eq:schrodinger}
\frac{d}{dt}|\varphi_k(t)\rangle\,=\,
-\frac{i}{\hbar}\Op{H}(\epsilon)|\varphi_k(t)\rangle\,,
\end{equation}
with an initial condition 
$|\varphi_k(0)\rangle=|\varphi_{ik}\rangle$.
Since $\Op{H}=\Op{H}(\epsilon)$ the state evolution will depend on the particular 
field. An alternative to Eq. (\ref{eq:schrodinger})
is the evolution equation for the unitary
transformation itself \cite{PK02}.

The method of optimization depends on the availability of the 
states of the system $|\varphi_k(t)\rangle$ at intermediate times $0 <t <T$.

Experimental realizations of OCT are typical examples where only initial and final 
knowledge of the states exist. For such cases 
feedback control and evolutionary methods are effective \cite{RZ00}. 
Such methods however require a large number of iterations 
to achieve convergence. A simulation of such a process
requires repeated propagation of the $N$ states by the 
Schr\"odinger equation. Thus they are computationally intensive.

Computationally more effective methods are based on  the knowledge
of the states $|\varphi_k(t)\rangle$ at intermediate times.
Additional constrains on the evolution are included that allow 
a modification of the field
at intermediate times consistent with the improvement
of the objective at the target time $T$.
Some examples are the local-in-time optimization method 
\cite{BKT93,Sugawara03},
the conjugate gradient search method \cite{KRGT89}, 
the Krotov method \cite{TKO92}, 
and the variational approach \cite{PDR88,ZBR98}.
A review of these common methods can be found in Ref. \cite{RZ00}.
In the present study the Krotov method has been adopted. A 
brief description of the alternative variational method is given
in Appendix \ref{ap:varapp}.

\subsection{Krotov method of optimization}

The Krotov method is utilized to derive an iterative 
algorithm to minimize a given functional that depends on
both final and intermediate times Cf. Ref. \cite{ST02}.

For convenience, the equations are stated using real functions:
$\alpha_{km}(t)\,\equiv\,{\rm Re}[\langle m|\varphi_k(t)\rangle]$
and
$\beta_{km}(t)\,\equiv\,{\rm Im}[\langle m|\varphi_k(t)\rangle]$.
The notation 
$\boldsymbol{\alpha}_k$ and $\boldsymbol{\beta}_k$ is used to describe
the $M$-dimensional vectors with components $\alpha_{km}$ 
and $\beta_{km}$. Using such a notation, the evolution equation 
(\ref{eq:schrodinger}) becomes:
\begin{eqnarray}\label{eq:evaandb}
\frac{d}{dt}\boldsymbol{\alpha}_k(t)\,&=&\,
\boldsymbol{\Omega}_R(t,\epsilon)\cdot\boldsymbol{\alpha}_k(t)\,-\,
\boldsymbol{\Omega}_I(t,\epsilon)\cdot\boldsymbol{\beta}_k(t)\,,\nonumber\\
\frac{d}{dt}\boldsymbol{\beta}_k(t)\,&=&\,
\boldsymbol{\Omega}_I(t,\epsilon)\cdot\boldsymbol{\alpha}_k(t)
\,+\,\boldsymbol{\Omega}_R(t,\epsilon)\cdot\boldsymbol{\beta}_k(t)
\,,
\end{eqnarray}
where $\boldsymbol{\Omega}_R$ and $\boldsymbol{\Omega}_I$ 
are real matrices with the corresponding components composed of
the real and imaginary parts of 
$\boldsymbol{\Omega}_{ij}=\langle i|(-i\Op{H}/\hbar)|j\rangle$ 
where $|i\rangle$ and $|j\rangle$ are states from the basis set $\{|m\rangle\}$.
The initial conditions are given by the vectors 
$\boldsymbol{\alpha}_k(0)$ and
$\boldsymbol{\beta}_k(0)$ with components composed of
the real and imaginary part
of the amplitudes $\langle m|\varphi_{ik}\rangle$.
$\boldsymbol{\alpha}_{fk}$ and 
$\boldsymbol{\beta}_{fk}$ denote the
vectors corresponding to the amplitudes $\langle m|\varphi_{fk}\rangle$.

The formalism  considers $\boldsymbol{\alpha}_k$,
$\boldsymbol{\beta}_k$, and the field $\epsilon$ to be independent variables.
A necessary consistency between them will be required
in the final step of the algorithm.  
The vectors 
$\boldsymbol{f}_{\alpha}$ and $\boldsymbol{f}_{\beta}$ 
constitute  the right hand side of Eq. (\ref{eq:evaandb}),
\begin{eqnarray}\label{eq:fandf}
\boldsymbol{f}_{\alpha}(t,\boldsymbol{\alpha}_k,
\boldsymbol{\beta}_k,\epsilon)
\,\equiv\,
\boldsymbol{\Omega}_R(t,\epsilon)\cdot\boldsymbol{\alpha}_k(t)\,-\,
\boldsymbol{\Omega}_I(t,\epsilon)\cdot\boldsymbol{\beta}_k(t)
\,,\nonumber\\
\boldsymbol{f}_{\beta}(t,\boldsymbol{\alpha}_k,
\boldsymbol{\beta}_k,\epsilon)
\,\equiv\,
\boldsymbol{\Omega}_I(t,\epsilon)\cdot\boldsymbol{\alpha}_k(t)\,+\,
\boldsymbol{\Omega}_R(t,\epsilon)\cdot\boldsymbol{\beta}_k(t)
\,.
\end{eqnarray}
The vectors 
$\boldsymbol{f}_\alpha$ ($\boldsymbol{f}_\beta$) 
are equal to the total time derivative of 
$\boldsymbol{\alpha}$ ($\boldsymbol{\beta}$) 
only when the state is consistent with the field through 
the evolution equation (\ref{eq:evaandb}).
The dependence of $\boldsymbol{\alpha}$ and $\boldsymbol{\beta}$ on
$t$ will be made explicit only when necessary.
An important property of the problems under study is that
$\boldsymbol{f}_{\alpha}$ and $\boldsymbol{f}_{\beta}$
are linear in the functions 
$\{\boldsymbol{\alpha},\boldsymbol{\beta}\}$
and $\epsilon$. This choice simplifies the optimization problem,
the non-linear case has been studied in Ref.  \cite{ST02}.

A ``process''
$w=w[t,\{\boldsymbol{\alpha},\boldsymbol{\beta}\},\epsilon]$ 
is defined as the set 
$\{\boldsymbol{\alpha},\boldsymbol{\beta}\}$ 
of  $N$ vectors $\boldsymbol{\alpha}_k$ and $N$ vectors 
$\boldsymbol{\beta}_k$ related to the field $\epsilon$ through the 
evolution equations with the initial conditions 
$\boldsymbol{\alpha}_k(0)$ and $\boldsymbol{\beta}_k(0)$.
A functional of the process can be defined:
\begin{equation}\label{eq:functional1}
J[w]\,=\,F(\{\boldsymbol{\alpha}(T),\boldsymbol{\beta}(T)\})\,+\,
\int_0^T g(\epsilon)\,dt\,.
\end{equation}
For the present applications $F$ can be any of the functionals
$F_{re}$, $F_{sm}$, $F_{ss}$ as introduced in the section 
\ref{se:implementation}.
The optimal field is  found by a minimization
of the functional $J$.
The integral term represents additional constrains 
originating from the evolution equation of the system. 
For simplicity only the case where $g$ is a  function
of the field $\epsilon$ is presented, but
a generalization to a more general case in which $g$ depends on 
$\boldsymbol{\alpha}(t)$ and $\boldsymbol{\beta}(t)$ is straightforward.
The particular dependence of $g$ on the field will be 
discussed later.

The main idea in the Krotov method is to introduce a new
functional that mixes the separated dependence on 
intermediate and final times in the original functional
(\ref{eq:functional1}). Using the new functional
it is possible to derive an iterative procedure that modifies
the field at intermediate times in a consistent way with
the minimization of $F$ at time $T$.
The new functional is defined as
\begin{equation}\label{eq:functional2}
L[w,\Phi]\,=\,G(\{\boldsymbol{\alpha}(T),\boldsymbol{\beta}(T)\})
\,-\,\Phi(0,\{\boldsymbol{\alpha}(0),\boldsymbol{\beta}(0)\})
\,-\,\int_0^T 
R(t,\{\boldsymbol{\alpha},\boldsymbol{\beta}\},\epsilon)\,dt\,,
\end{equation}
where,
\begin{equation}\label{eq:gkrotov}
G(\{\boldsymbol{\alpha}(T),\boldsymbol{\beta}(T)\})\,=\,
F(\{\boldsymbol{\alpha}(T),\boldsymbol{\beta}(T)\})\,+\,
\Phi(T,\{\boldsymbol{\alpha}(T),\boldsymbol{\beta}(T)\})\,,
\end{equation}
and
\begin{eqnarray}\label{eq:rkrotov}
R(t,\{\boldsymbol{\alpha},\boldsymbol{\beta}\},\epsilon)\,&=&\,
-g(\epsilon)
\,+\,\frac{\partial\Phi}{\partial t}
(t,\{\boldsymbol{\alpha},\boldsymbol{\beta}\}) \nonumber\\
\,&+&\,\sum_{k=1}^{N}\,
\boldsymbol{\frac{\partial\Phi}{\partial\alpha_k}}
(t,\{\boldsymbol{\alpha},\boldsymbol{\beta}\})
\cdot
\boldsymbol{f}_{\alpha}
(t,\boldsymbol{\alpha}_k,\boldsymbol{\beta}_k,\epsilon)\nonumber\\
\,&+&\,\sum_{k=1}^{N}\,
\boldsymbol{\frac{\partial\Phi}{\partial\beta_k}}
(t,\{\boldsymbol{\alpha},\boldsymbol{\beta}\})
\cdot
\boldsymbol{f}_{\beta}
(t,\boldsymbol{\alpha}_k,\boldsymbol{\beta}_k,\epsilon)\,.
\end{eqnarray}
$\Phi(t,\{\boldsymbol{\alpha},\boldsymbol{\beta}\})$ 
denotes an arbitrary continuously differentiable function.
The partial derivatives  of $\Phi$,
$\boldsymbol{\frac{\partial\Phi}{\partial \alpha_k}}$ and
$\boldsymbol{\frac{\partial\Phi}{\partial \beta_k}}$,
form a vector with $m$ components.
In the following $t$, $\boldsymbol{\alpha}$ and
$\boldsymbol{\beta}$ are considered to be independent variables
in $\Phi$.

When 
$\{\boldsymbol{\alpha},\boldsymbol{\beta}\}$ 
and the field are related by Eq. (\ref{eq:evaandb}), 
$R$ can be written as $R=-g+d\Phi/dt$. 
Introducing this result in Eq. (\ref{eq:functional2}),
it can be shown \cite{ST02} that for any scalar function $\Phi$ and
any process $w$, $L[w,\Phi]=J[w]$.
Then the minimization of $J$ is completely 
equivalent to the minimization of $L$.

\subsubsection{Iterative algorithm to minimize $L$}

The advantage of the definition of
the functional $L[w]$ is the complete freedom in the 
choice of $\Phi$. 
This property is used to derive from an arbitrary
process 
$w^{(0)}[t,\{\boldsymbol{\alpha}^{(0)},
\boldsymbol{\beta}^{(0)}\},\epsilon^{(0)}]$
a new process 
$w^{(1)}[t,\{\boldsymbol{\alpha}^{(1)},
\boldsymbol{\beta}^{(1)}\},\epsilon^{(1)}]$
such that $L[w^{(1)},\Phi]\leq L[w^{(0)},\Phi]$.
The procedure can be summarized as follows:

\begin{itemize}
\item{(i) $\Phi$ is constructed so that the functional $L[w^{(0)}]$ is 
a maximum with respect to any possible choice of the set
$\{\boldsymbol{\alpha},\boldsymbol{\beta}\}$. 
This condition gives a complete freedom to change $\epsilon$.
The related changes of the states are consistent
with the system evolution, Eq. (\ref{eq:evaandb}), 
and therefore, will not interfere with the  the minimization of $L$.}

\item{(ii) A new field $\epsilon^{(1)}$ is derived with the 
condition of maximizing $R$, decreasing then the value
of $L$ with respect to the process $w^{(0)}$. 
In this step the consistency between the new
field and the new states of the system 
$\{\boldsymbol{\alpha}^{(1)},\boldsymbol{\beta}^{(1)}\}$ 
must be maintained.}
\end{itemize}

The new field $\epsilon^{(1)}$ becomes the starting point of a new 
iteration, and steps (i) and (ii) are repeated until the 
desired convergence is achieved.

\subsubsection{The linear problem: construction of
$\Phi$ to first order}

The difficult task in the Krotov method is the construction of 
$\Phi$ so that $L$ is maximum for 
$\{\boldsymbol{\alpha}^{(0)},\boldsymbol{\beta}^{(0)}\}$. 
The maximum condition on $L$ is equivalent to imposing a maximum 
on $G$ and a minimum on $R$. 
However, in some cases the maximum and minimum conditions can 
be relaxed to extreme conditions for $G$ and $R$, which simplifies 
the problem.

The extreme conditions for $R$ with respect to 
$\{\boldsymbol{\alpha}^{(0)},\boldsymbol{\beta}^{(0)}\}$ 
are given by 
\begin{eqnarray}\label{eq:extr}
\boldsymbol{\frac{\partial R}{\partial\alpha_k}}
(t,\{\boldsymbol{\alpha}^{(0)},\boldsymbol{\beta}^{(0)}\},\epsilon^{(0)})
&=&0\,,\nonumber\\
\boldsymbol{\frac{\partial R}{\partial\beta_k}}
(t,\{\boldsymbol{\alpha}^{(0)},\boldsymbol{\beta}^{(0)}\},\epsilon^{(0)})
&=&0\,.
\end{eqnarray}
The following vectors are introduced
\begin{eqnarray}\label{eq:gandd}
\boldsymbol{\gamma}_k(t)\,&=&\,
\boldsymbol{\frac{\partial\Phi}{\partial\alpha_k}}
(t,\{\boldsymbol{\alpha}^{(0)},\boldsymbol{\beta}^{(0)}\})\,,\nonumber\\
\boldsymbol{\delta}_k(t)\,&=&\,
\boldsymbol{\frac{\partial\Phi}{\partial\beta_k}}
(t,\{\boldsymbol{\alpha}^{(0)},\boldsymbol{\beta}^{(0)}\})\,.
\end{eqnarray}
$\boldsymbol{\gamma}_k$ and $\boldsymbol{\delta}_k$ 
are only functions of $t$, as the
partial derivatives are evaluated in the specific set 
$\{\boldsymbol{\alpha}^{(0)},\boldsymbol{\beta}^{(0)}\}$. 
Using Eq. (\ref{eq:evaandb}) the extreme conditions
can be written as
\begin{eqnarray}\label{eq:evgandd}
\frac{d}{dt}\,\boldsymbol{\gamma}_k(t) &=&\,
-\boldsymbol{\Omega}_R^T(t,\epsilon^{(0)})\cdot
\boldsymbol{\gamma}_k(t)
\,-\,\boldsymbol{\Omega}_I^T(t,\epsilon^{(0)})\cdot
\boldsymbol{\delta}_k(t)\,,\nonumber\\
\frac{d}{dt}\,\boldsymbol{\delta}_k(t) &=&\,
\boldsymbol{\Omega}_I^T(t,\epsilon^{(0)})\cdot
\boldsymbol{\gamma}_k(t)
\,-\,\boldsymbol{\Omega}_R^T(t,\epsilon^{(0)})\cdot
\boldsymbol{\delta}_k(t)\,,
\end{eqnarray}
where $\boldsymbol{\Omega}^T$ denotes the transpose of the matrix 
$\boldsymbol{\Omega}$.
The extreme conditions for $G$ are
\begin{eqnarray}\label{eq:extg}
\boldsymbol{\frac{\partial G}{\partial\alpha_k(T)}}
(\{\boldsymbol{\alpha}^{(0)}(T),
\boldsymbol{\beta}^{(0)}(T)\})\,&=&\,0\,,\nonumber\\
\boldsymbol{\frac{\partial G}{\partial\beta_k(T)}}
(\{\boldsymbol{\alpha}^{(0)}(T),
\boldsymbol{\beta}^{(0)}(T)\})\,&=&\,0\,.
\end{eqnarray}
Using Eq. (\ref{eq:gkrotov}) and Eq. (\ref{eq:gandd}) 
\begin{eqnarray}\label{eq:cogandd}
\boldsymbol{\gamma}_k(T)\,&=&\,
-\boldsymbol{\frac{\partial F}{\partial\alpha_k(T)}}
(\{\boldsymbol{\alpha}{(0)}(T),\boldsymbol{\beta}^{(0)}(T)\})\,,
\nonumber\\
\boldsymbol{\delta}_k(T)\,&=&\,
-\boldsymbol{\frac{\partial F}{\partial\beta_k(T)}}
(\{\boldsymbol{\alpha}^{(0)}(T),\boldsymbol{\beta}^{(0)}(T)\})\,.
\end{eqnarray}
The above conditions at time $T$, together with Eq. (\ref{eq:evgandd})
determine completely the set  
$\{\boldsymbol{\gamma}(t),\boldsymbol{\delta}(t)\}$. 
As they are defined as the partial
derivative of $\Phi$ with respect to $\alpha_{km}$ and $\beta_{km}$,
$\Phi$ is expanded to first order (denoted as $\Phi^\star$),
\begin{equation}\label{eq:Phi0}
\Phi^\star(t,\{\boldsymbol{\alpha},\boldsymbol{\beta}\})\,=\,
\sum_{k=1}^{N}\,\left[
\boldsymbol{\gamma}_k(t)\cdot\boldsymbol{\alpha}_k(t)\,+\,
\boldsymbol{\delta}_k(t)\cdot\boldsymbol{\beta}_k(t)\right]\,.
\end{equation}
By employing $\Phi^\star$, the functions 
$G^\star$ and $R^\star$ can also be constructed to first order using 
Eq. (\ref{eq:gkrotov}) and (\ref{eq:rkrotov}) respectively.
This completes the first step in the iterative algorithm.

To accomplish the second step $R$ is maximized with respect to
the field. Again in some cases the maximum condition
can be relaxed  to the extreme condition 
$\partial R/\partial \epsilon=0$.
Using the expression for $R^\star$ leads to
\begin{eqnarray}\label{eq:newe1}
\frac{\partial g}{\partial\epsilon}(\epsilon^{(1)})
\,&=&\,
\sum_{k=1}^{N}\,\boldsymbol{\gamma}_k(t)\cdot
\frac{\partial \boldsymbol{f}_{\alpha}}{\partial\epsilon}
(t,\boldsymbol{\alpha}_k^{(1)},
\boldsymbol{\beta}_k^{(1)},\epsilon^{(1)})\nonumber\\
\,&+&\,
\sum_{k=1}^{N}\,\boldsymbol{\delta}_k(t)\cdot
\frac{\partial \boldsymbol{f}_{\beta}}{\partial\epsilon}
(t,\boldsymbol{\alpha}_k^{(1)},
\boldsymbol{\beta}_k^{(1)},\epsilon^{(1)})\,.
\end{eqnarray}
Eq. (\ref{eq:newe1}) is used to derive the new field $\epsilon^{(1)}$
in each iteration. This equation must be solved in a consistent
way with Eq. (\ref{eq:evaandb}) describing the system dynamics.

Due to the use of extreme instead of maximum or minimum conditions,
it must be checked that the new process, $w^{(1)}$, 
improves the original objective in each iteration
$J[w^{(0)}]-J[w^{(1)}]\geq 0$:
\begin{equation}
J[w^{(0)}]-J[w^{(1)}]\,=\,
L[w^{(0)},\Phi^\star]-L[w^{(1)},\Phi^\star]\,=\,\Delta_1
\,+\,\int_0^T\Delta_2(t)\,dt\,.
\end{equation}
where,
\begin{equation}\label{eq:delta1}
\Delta_1\,=\,
G^\star(\{\boldsymbol{\alpha}^{(0)}(T),
\boldsymbol{\beta}^{(0)}(T)\})\,-\,
G^\star(\{\boldsymbol{\alpha}^{(1)}(T),
\boldsymbol{\beta}^{(1)}(T)\})\,.
\end{equation}
\begin{equation}\label{eq:delta2}
\Delta_2(t)\,=\,
R^\star(t,\{\boldsymbol{\alpha}^{(1)},
\boldsymbol{\beta}^{(1)}\},\epsilon^{(1)})\,-\,
R^\star(t,\{\boldsymbol{\alpha}^{(1)},
\boldsymbol{\beta}^{(1)}\},\epsilon^{(0)})\,.
\end{equation}
The above relation is obtained when $\boldsymbol{f}_{\alpha}$ and 
$\boldsymbol{f}_{\beta}$ are linear in
$\{\boldsymbol{\alpha},\boldsymbol{\beta}\}$
\begin{equation}\label{eq:rforep0}
R^\star(t,\{\boldsymbol{\alpha},\boldsymbol{\beta}\},
\epsilon^{(0)})\,=\,-\,g(t,\epsilon^{(0)})\,,
\end{equation}
for any value of $\{\boldsymbol{\alpha},\boldsymbol{\beta}\}$.

A sufficient condition for $J[w^{(0)}]-J[w^{(1)}]\geq 0$
is $\Delta_1,\Delta_2(t)\geq 0$.
$\Delta_1$ depends on the choice $F$ and
$\Delta_2(t)$ on the choice of $g$ so that each case
must be analyzed separately.
These conditions imply that the Krotov iterative algorithm
convergence monotonically to the final objective.

\subsubsection{Dependence on $F$}

The dependence of $G^\star$ on $F$ can be made explicit
by introducing $\Phi^\star$ and using Eq. (\ref{eq:Phi0}) and
Eq. (\ref{eq:gkrotov}), 
\begin{eqnarray}
G^\star(\{\boldsymbol{\alpha}(T),
\boldsymbol{\beta}(T)\})\,&=&\,
F(\{\boldsymbol{\alpha}(T),\boldsymbol{\beta}(T)\})\nonumber\\
\,&-&\,\sum_{k=1}^{N}\,
\boldsymbol{\frac{\partial F}{\partial \alpha_k(T)}}
\cdot\boldsymbol{\alpha}_k(T)\,
\,+\,\boldsymbol{\frac{\partial F}{\partial \beta_k(T)}}
\cdot\boldsymbol{\beta}_k(T)\,.
\end{eqnarray}
When $F$ is linear in 
$\{\boldsymbol{\alpha},\boldsymbol{\beta}\}$ 
$G^\star\equiv 0$ and then $\Delta_1\equiv 0$. 
In this case all the improvement towards the original objective
in the iteration is due to the term $\Delta_2(t)$.
When $F$ is non linear in $\{\boldsymbol{\alpha},\boldsymbol{\beta}\}$
the condition $\Delta_1\leq 0$ must be checked in each case.

An additional difficulty is that the conditions (\ref{eq:gandd})
for $\boldsymbol{\gamma}$ and $\boldsymbol{\delta}$
depend on the particular choice of $F$.
In all the cases under study ($F_{re}$, $F_{sm}$ and $F_{ss}$) 
\begin{eqnarray}\label{eq:gamT}
\boldsymbol{\gamma}_k(T)\,&=&\,c_k\,\boldsymbol{\alpha}_{fk}(T)\,,
\nonumber\\
\boldsymbol{\delta}_k(T)\,&=&\,d_k\,\boldsymbol{\beta}_{fk}(T)\,,
\end{eqnarray}
where the coefficients $c_k$ and $d_k$ depend on 
the sets 
$\{\boldsymbol{\alpha}^{(0)}(T),\boldsymbol{\beta}^{(0)}(T)\}$ 
and $\{\boldsymbol{\alpha}_f,\boldsymbol{\beta}_f\}$.
Defining the vectors 
$\tilde{\boldsymbol{\gamma}}_k=c_k^{-1}\,\boldsymbol{\gamma}_k$ 
and
$\tilde{\boldsymbol{\delta}}_k=d_k^{-1}\,\boldsymbol{\delta}_k$, 
the conditions (\ref{eq:cogandd}) for all the cases 
under consideration are: 
\begin{eqnarray}\label{eq:tilgamT}
\tilde{\boldsymbol{\gamma}}_k(T)\,&=&\,
\boldsymbol{\alpha}_{fk}(T)\,,\nonumber\\
\tilde{\boldsymbol{\delta}}_k(T)\,&=&\,
\boldsymbol{\beta}_{fk}(T)\,.
\end{eqnarray}
Their evolution is given by Eq. (\ref{eq:evgandd}).
Changing $\boldsymbol{\gamma}$ and 
$\boldsymbol{\delta}$ to $\tilde{\boldsymbol{\gamma}}$ and
$\tilde{\boldsymbol{\delta}}$ Eq. (\ref{eq:newe1}) can be written as
\begin{eqnarray}\label{eq:newe2}
\frac{\partial g}{\partial\epsilon}(\epsilon^{(1)})
\,&=&\,
\sum_{k=1}^{N}\,c_k\,\tilde{\boldsymbol{\gamma}}_k(t)\cdot
\frac{\partial {\boldsymbol f}_{\alpha}}{\partial\epsilon}
(t,\boldsymbol{\alpha}_k^{(1)},\boldsymbol{\beta}_k^{(1)},
\epsilon^{(1)})\nonumber\\
\,&+&\,
\sum_{k=1}^{N}\,d_k\,\tilde{\boldsymbol{\delta}}_k(t)\cdot
\frac{\partial \boldsymbol{f}_{\beta}}{\partial\epsilon}
(t,\boldsymbol{\alpha}_k^{(1)},
\boldsymbol{\beta}_k^{(1)},\epsilon^{(1)})\,.
\end{eqnarray}
The different choices of $F$ imply  different coefficients
($c_k$ and $d_k$) and a possible different set of initial 
$|\varphi_{ik}\rangle$
and final $|\varphi_{fk}\rangle$ states. Nevertheless, 
the iterative procedure is identical in all the cases.

\subsubsection{Dependence on $g(\epsilon)$}

A delicate point is the choice of the function
$g(\epsilon)$ in $J[w]$. The time integral in the functional 
should be bounded from below,
otherwise the the additional constraint will dominate over
the original objective $F$ in the functional $J$.
In addition  $\Delta_2(t)\geq 0$ is required in order to 
guarantee the  monotonic convergence of the optimization method.

A consequence of the linear dependence of 
$\boldsymbol{f}_\alpha$, $\boldsymbol{f}_\beta$
and Eq. (\ref{eq:newe2}) is that the function $R^\star$ 
for the new process $w^{(1)}$ has the simple form
\begin{equation}
R^\star(t,\{\boldsymbol{\alpha}^{(1)},
\boldsymbol{\beta}^{(1)}\},\epsilon^{(1)})
\,=\,-g(\epsilon)\,+\,\left(\epsilon^{(1)}-\epsilon^{(0)}\right)\,
\frac{\partial g}{\partial \epsilon}(\epsilon^{(1)})\,.
\end{equation}
Using this expression together with Eq. (\ref{eq:rforep0}) leads to
\begin{equation}\label{eq:condd2}
\Delta_2(t)\,=\,-g(\epsilon^{(1)})+g(\epsilon^{(0)})
\,+\,\left(\epsilon^{(1)}-\epsilon^{(0)}\right)
\frac{\partial g}{\partial\epsilon}(\epsilon^{(1)})\,,
\end{equation}
A choice of $g(\epsilon)$ fulfilling the previous requirements 
is
\begin{equation}\label{eq:gchoice}
g(\epsilon)\,=\,\lambda(t)\,
\left[\epsilon(t)\,-\,\tilde{\epsilon}(t)\right]^2\,,
\end{equation}
where $\tilde{\epsilon}$ is a reference field and
$\lambda(t)$ is a positive function of $t$. Using
Eq. (\ref{eq:condd2}) and for any field $\tilde{\epsilon}$
\begin{equation}
\Delta_2(t)\,=\,\lambda(t)\,
\left[\Delta\epsilon(t)\right]^2\,\,\geq\,0,
\end{equation}
where $\Delta\epsilon(t)\equiv\epsilon^{(1)}(t)-\epsilon^{(0)}(t)$.
The method therefore will converge monotonically.
Using Eq. (\ref{eq:newe2}) and Eq. (\ref{eq:gchoice})
the field in the new iteration becomes:
\begin{eqnarray}\label{eq:newe3}
\epsilon^{(1)}(t)
\,&=&\,
\tilde{\epsilon}(t)
\,+\,
\frac{1}{2\lambda(t)}\,
\sum_{k=1}^{N}\,
\left\{
c_k\,\tilde{\boldsymbol{\gamma}}_k(t)\cdot
\frac{\partial \boldsymbol{f}_{\alpha}}{\partial\epsilon}
(t,\boldsymbol{\alpha}_k^{(1)},\boldsymbol{\beta}_k^{(1)},
\epsilon^{(1)})\right.\nonumber\\
\,&+&\,
\left. d_k\,\tilde{\boldsymbol{\delta}}_k(t)\cdot
\frac{\partial \boldsymbol{f}_{\beta}}{\partial\epsilon}
(t,\boldsymbol{\alpha}_k^{(1)},\boldsymbol{\beta}_k^{(1)},
\epsilon^{(1)})\right\}\,.
\end{eqnarray}
The result of the iterative algorithm depends strongly on the
choices of the reference field $\tilde{\epsilon}$ and on the 
function $\lambda(t)$.


Two possible choices of $\tilde{\epsilon}$ are analyzed.
The first, $\tilde{\epsilon}=0$, is the one commonly used in OCT 
applications \cite{RZ00}. In this case, the additional 
constraint in $J[w]$ has the physical meaning that the total
energy of the field in the time interval $[0,T]$ is 
limited.
This however presents a problem when the iterative
procedure reaches the optimal field. The iterative method
is found to reduce the total objective $J$
by reducing the total pulse energy, slowing and even
spoiling the convergence to the original objective $F$.
The usual remedy is to stop the iterative algorithm 
before this difficulty is reached. However, such a  procedure
could prevent the optimization algorithm from obtaining 
high fidelity. 

A different possibility is $\tilde{\epsilon}=\epsilon^{(0)}$ 
can avoid this problem \cite{RB01,BKT01,ST02}.
In this iterative algorithm $\epsilon^{(0)}$ must be interpreted 
as the field in the previous iteration. 
Now the additional constraint in $J[w]$ has the
physical interpretation that the change of the pulse energy 
in each iteration is limited. When the iterative procedure
approaches the optimal solution the change
in the field vanishes. Therefore, the convergence 
to the original objective is guaranteed.
In the rest of the study  $\tilde{\epsilon}=\epsilon^{(0)}$ was chosen.


The function $\lambda(t)$ introduces the shape function $s(t)$ i.e.
$\lambda(t)=\lambda_0/s(t)$. The purpose of $s(t)$ is to  turn the field on 
and off smoothly  at the boundaries of the interval \cite{SV99}.
$\lambda_0$ is a scaling parameter which determines the optimization strategy.
When $\lambda_0$ is small 
the additional constraint on the field in the functional becomes insignificant, 
resulting in large modifications in the field in each
iteration. This is equivalent to a bold search strategy 
where large excursions in the functional space of the field
take place with each iteration.
Large values of $\lambda_0$ imply small
modifications in the field in each iteration, slowing the
convergence process. 
Using large values of $\lambda_0$ is a conservative search strategy
which is advantageous when a good initial guess field
can be found. 
A possible mixed strategy is to initially use 
a bold optimization with small values of $\lambda_0$.
This leads to a  guess field for a new optimization with a large
value of $\lambda_0$  \cite{HMV02}.

\subsection{Application to the functionals $F_{re}$, $F_{sm}$ and $F_{ss}$}

Based on the derivation of the Krotov method it is possible
to connect directly the minimization $F_{re}$, $F_{sm}$ and $F_{ss}$
to the correction to the field.
Eq. (\ref{eq:evgandd}) corresponds to the evolution of 
a set of states $\{|\chi_k(t)\rangle\}$,
\begin{equation}\label{eq:evchi}
\frac{d}{dt}|\chi_k(t)\rangle\,=\,
-\frac{i}{\hbar}\,\Op{H}^\dagger(\epsilon^{(0)})
\,|\chi_k(t)\rangle\,,
\end{equation}
with the conditions (\ref{eq:tilgamT}), 
$|\chi_k(T)\rangle=|\varphi_{fk}\rangle$. 
The formal solution of the equation is given by
$|\chi_k(t)\rangle=\Op{U}(t,T;\epsilon^{(0)})\Op{O}|\varphi_{ik}\rangle$.
Using Eq. (\ref{eq:newe3}) the correction to the field
in each iteration becomes:
\begin{equation}\label{eq:newe}
\Delta\epsilon(t)\,=\,
-\frac{s(t)}{\lambda_0\,\hbar}\,{\rm Im}\left[
\sum_{k=1}^{N}\,a_k(\epsilon^{(0)})\,
\la \varphi_{ik}|\Op{O}^\dagger\Op{U}^\dagger(t,T;\epsilon^{(0)})
\,\Op{\mu}\,
\Op{U}(t,0;\epsilon^{(1)})|\varphi_{ik}\ra
\right]\,.
\end{equation}
The coefficients $a_k$, will depend on the particular choice 
of the functional $F$ and are related to $c_k$ and $d_k$ defined
in Eq. (\ref{eq:gandd}).
For $F_{re}$ and $F_{sm}$, the states $\{|\varphi_{ik}\rangle\}$ 
denote an orthonormal basis $\{|n\rangle\}$ of the relevant 
subspace $\cal N$. For $F_{re}$ the coefficients are $a_{n,re}=1/2$,
and as this functional is linear on the states $\Delta_{1,re}=0$.
For $F_{sm}$ the coefficients are
\begin{equation}\label{eq:asm}
a_{n,sm}\,=\,\sum_{n^\prime=1}^{N}\,
\la n^\prime|\Op{U}(0,T;\epsilon^{(0)})\,
\Op{O}|n^\prime\rangle\,,
\end{equation}
and then are equal for all the states in the basis of ${\cal N}$.
In addition,
\begin{equation}
\Delta_{1,sm}\,=\,
\left|\sum_{n=1}^{N}\, 
\la n|
\left(\Op{U}(T,0;\epsilon^{(0)})-\Op{U}(T,0;\epsilon^{(1)})\right)
\Op{O}\,|n\, \ra
\right|^2\,.
\end{equation}
Therefore, $\Delta_{1,sm}\geq 0$.
For the $F_{ss}$ functional the set $\{|\varphi_{ik}\rangle\}$ for which the states
are denoted by $\{|l\rangle\}$ the coefficients
$a_l$ are
\begin{equation}\label{eq:ass}
a_{l,ss}\,=\,
\la l|\Op{U}(0,T;\epsilon^{(0)})
\Op{O}|l\ra\,,
\end{equation}
depending on the index $l$ corresponding to each state. 
In this case 
\begin{equation}
\Delta_{1,ss}\,=\,
\sum_{l=1}^{N}\,\left| 
\la l|
\left(\Op{U}(T,0;\epsilon^{(0)})-\Op{U}(T,0;\epsilon^{(1)})\right)
\Op{O}
|l \ra
\right|^2\,,
\end{equation}
and then $\Delta_{1,ss}\geq 0$.

The results $\Delta_1\geq 0$ and $\Delta_2(t)\geq 0$
guarantee the monotonic convergence of the iterative
algorithm based on the Krotov method for the three functionals.


\subsection{The optimal field}

The optimal field has the property that
the field correction in the next iteration
Eq. (\ref{eq:newe}) should vanish.
Defining this correction as:
\begin{equation}\label{eq:ceps}
C(t;\epsilon)\,=\,{\rm Im}\left[
\sum_{k=1}^{N}\,a_k(\epsilon^{(0)})\,
\la \varphi_{ik}|\Op{O}^\dagger\Op{U}^\dagger(t,T;\epsilon)
\,\Op{\mu}\,
\Op{U}(t,0;\epsilon)|\varphi_{ik}\ra
\right]\,,
\end{equation}
where $\bar{\epsilon}$ is an arbitrary solution for which
$C(t;\bar{\epsilon})\equiv0$.

The first question to be addressed is whether any optimal field,
defined by Eq. (\ref{eq:condition}), is a possible solution of the
iterative algorithm.
$\bar{\epsilon}_{opt}$ denotes a field that 
generates the target unitary transformation up to a global
phase, $\Op{U}(T,0;\bar{\epsilon}_{opt})=e^{-i\bar{\phi}}\,\Op{O}$.
Using the relation
\begin{equation}\label{eq:uandu}
\Op{U}(t,0;\epsilon)=\Op{U}(t,T;\epsilon)\,\Op{U}(T,0;\epsilon)\,.
\end{equation}
In addition, the relation 
$|\Psi_k(t)\rangle=\Op{U}(t,T;\bar{\epsilon}_{opt})
\,\Op{O}|\varphi_{ik}\rangle$ implying that
the term $\langle\Psi_k(t)|\Op{\mu}|\Psi_k(t)\rangle$ is real,
simplifies Eq. (\ref{eq:ceps}) to:
\begin{equation}\label{eq:cepsop}
C(t;\bar{\epsilon}_{opt})\,=\,
\sum_{k=1}^{N}\,
\langle\Psi_k(t)|\,\Op{\mu}\,|\Psi_k(t)\rangle\,
{\rm Im}\left[\,a_k(\bar{\epsilon}_{opt})\,e^{-i\bar{\phi}}\,
\right]\,,
\end{equation}

Using Eq. (\ref{eq:asm}) for the functional $F_{sm}$ leads to
$a_{n,sm}=N\exp{(i\bar{\phi})}$ in Eq. (\ref{eq:ceps}). 
A similar result is found for the functional $F_{ss}$, 
$a_{l,ss}=\exp{(i\bar{\phi})}$, given by Eq. (\ref{eq:ass}).
Therefore any field generating the target unitary transformation
is a possible solution of the iterative algorithm based on 
any of the functionals $F_{sm}$ and $F_{ss}$. This result 
does not imply that when initializing the different iteration schemes
with the same guess  field the same solution will be obtained.

The analysis is more complex for the functional $F_{re}$.
The coefficients are now real $a_{n,re}=1/2$,  
and are independent of the state index.  This leads to 
$C_{re}={\rm Im}[\exp{(-i\bar{\phi})}]\sum_{n=1}^{N}
\langle\Psi_k(t)|\,\Op{\mu}\,|\Psi_k(t)\rangle$.
The sum is generally different from zero and the solutions
to the algorithm are fields with a phase term 
$\exp{(-i\bar{\phi})}=\pm 1$. 
Only the case $+1$ minimize the original functional $F_{re}$,
but the relaxation to extreme conditions in the Krotov method
allows to obtain other physically valid solutions.
In the special case in which the unitary transformation is imposed on all
the Hilbert space ($N=M$), any optimal
field is a possible solution regardless of the global phase. The reason is
that the sum in $C_{re}$ is zero since $\Op{\mu}$ is a traceless operator. 

The phase sensitivity of the functional
$F_{re}$ can be demonstrated in the state-to-state
optimization.
The iterative algorithm in this case will converge to a 
field that drives the system to the final state 
$+|\varphi_f\rangle$ or $-|\varphi_f\rangle$, while 
the optimization of $F_{sm}$ or $F_{ss}$ will converge
to the final state up to an arbitrary global phase.
There is no a priori advantage however to any of the three
functionals in the convergence rate or in the simplicity
of the solution. The solutions are physically equivalents
since they differ only in a global phase.

In addition to the desired optimal fields, the algorithm could also
generate spurious solutions. A possible example is the functional
$F_{sm}$ employed to implement a unitary transformation
$\Op{O}_D$ with a matrix representation diagonal in the 
basis of the free Hamiltonian eigenstates $|e_n\rangle$.
In such a case $C(t,\epsilon=0)$ is proportional to the
diagonal matrix elements 
$\langle e_n|\Op{\mu}|e_n\rangle$.
When these matrix elements are zero,
$\epsilon=0$ is a solution of the iterative algorithm, 
but it does not implement the desired unitary transformation.
A simple remedy to overcome this difficulty is to use a different 
initial guess to start the algorithm.


\subsection{Discrete implementation of the optimization algorithm}

A numerical solution of the iterative optimization
algorithm requires a discretization scheme for the time axes.
The correction to the field
$\Delta\epsilon$ is implicit and appears on
both sides of Eq. (\ref{eq:newe}). To implement the procedure,
two interleaved grid points in time were used. 
The first grid was used to
propagate the  states. The second grid was used to evaluate
the field.
The grid describing the states has $N_t+1$ points separated by
$\Delta t=T/N_t$, from $t=0$ to $t=T$.
The grid representing the field has $N_t$ points separated by $\Delta t$
and starting at $t=\Delta t/2$. 
The initial set of states $|\varphi_{ik}\rangle$ was used for
the target unitary transformation $\Op{O}$ optimization
with the functionals $F_{re}$, $F_{sm}$ or $F_{ss}$.
The numerical implementation of the algorithm 
follows:
\begin{itemize}
\item{(i) Using an initial guess field $\epsilon^{(0)}$, the states
$\varphi_{fk}$ are propagated in reverse from $t=T$ to $t=0$
to determine  $\Op{U}(t,T;\epsilon^{(0)})\Op{O}|\varphi_{ik}\rangle$
on the time grid of states.}

\item{(ii) The new field is determined in the interleaved grid point
$t=\Delta t/2$ using the approximation
\begin{equation}\label{eq:neweaprox}
\Delta\epsilon(\Delta t/2)\,\approx\,
-\frac{s(\Delta t/2)}{\lambda_0\,\hbar}\,{\rm Im}\left[
\sum_{k=1}^{N}\,a_k(\epsilon^{(0)})\,
\la \varphi_{ik}|\Op{O}^\dagger\Op{U}^\dagger(0,T;\epsilon^{(0)})
\,\Op{\mu}\,
\Op{U}(0,0;\epsilon^{(1)})|\varphi_{ik}\ra
\right]\,.
\end{equation}
Notice that 
$\Op{U}(0,0;\epsilon^{(1)})|\varphi_{ik}\rangle=|\varphi_{ik}\rangle$.
Then the new field in the first field time grid point is
obtained,
$\epsilon^{(1)}(\Delta t/2)=\epsilon^{(0)}+\Delta\epsilon(\Delta t/2)$ 
and used to propagate $|\varphi_{ik}(t=0)\rangle$
to the next state grid point $t=\Delta t$.
The same process is used to obtain the new field
$\epsilon^{(1)}$ in the next field time grid point 
$t=\Delta t+\Delta t/2$, evaluating the correction
with the already know states in the state grid point
$t=\Delta t$.
The process is repeated to obtain $\epsilon^{(1)}$ in all 
the field time grid points.}
\item{
(iii) The new field $\epsilon^{(1)}$ is used as input
to the new iteration ($\epsilon^{(0)}=\epsilon^{(1)})$
and the process is repeated until the required convergence
is achieved.}
\end{itemize}

More elaborate methods to deal with the implicit time dependence
of Eq. (\ref{eq:newe}) have been developed. For example, approximating
the dynamics in between grid points by
the free evolution with $\Op{H}_0$ \cite{ZBR98}.
The simple procedure, which is able to keep the monotonic behavior 
of the optimization method was found sufficient.

The present implementation is based on a forward time propagation. 
Using the same formalism, the optimization can be accomplished 
also by a backward time propagation.
It is also possible to combine both cases, and to perform
the optimization in the
forward and backward propagations \cite{ZBR98,MT03}. 
In the current studies, these other procedures were found to be 
inferior, slowing down the convergence rate.


\section{The Fourier transform example 
in a molecular model}\label{se:qubits}

\begin{figure}[h]
\vspace{1.2cm}
\hspace{0.16\textwidth}
\psfig{figure=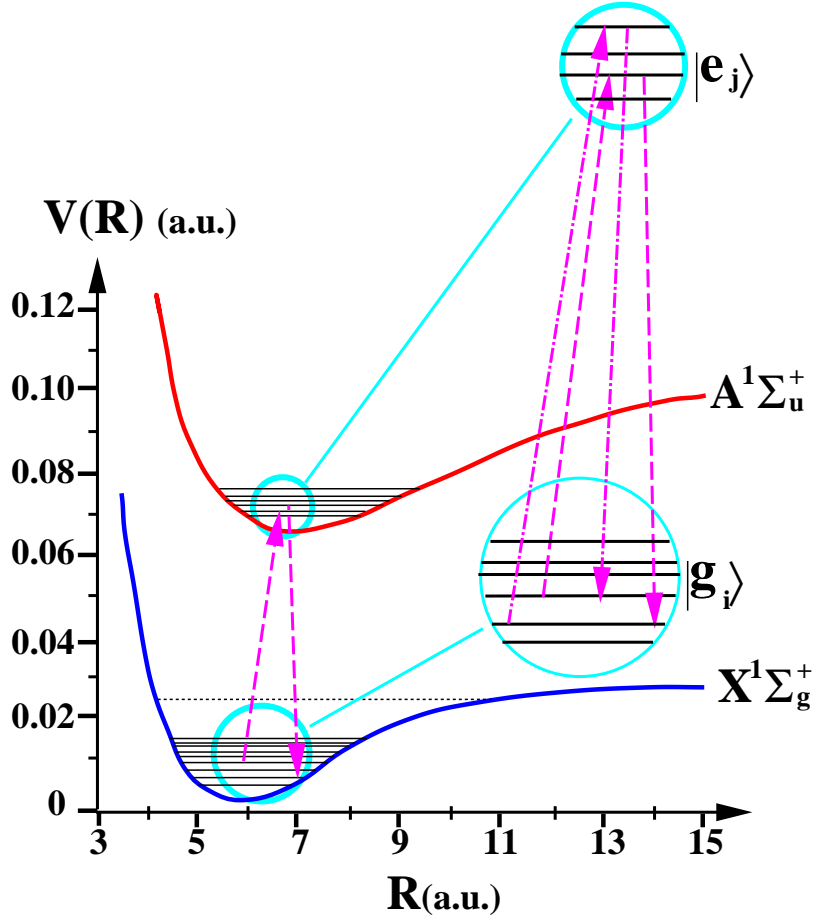,width=0.65\textwidth}
\vspace{.3cm}
\caption{Schematic representation of a molecular  model based on 
the vibrational levels in the $X^1\Sigma^+_g$ (lower) and
$A^1\Sigma^+_u$ (upper) electronic surfaces of
the molecule Na$_2$.
Atomic units are chosen $\hbar=1$. $R$ denotes the internuclear distance.
The arrows indicate two of the possible 
transitions induced by the driving field between arbitrary levels
in the lower and upper surfaces. On the right is a magnified view
of some of the energy levels involved and transitions between them.} 
\label{fig:schme}
\end{figure}

As an illustration the implementation
of $Q$ qubit  Fourier transform in a two electronic surfaces 
molecular model was studied. Fig. \ref{fig:schme} shows a schematic
view of a model based on the electronic manifolds of Na$_2$. 

The Hamiltonian of the system describes a ground and excited electronic 
potential energy surface coupled by a transition dipole operator:
\begin{equation}\label{eq:hamhad}
\Op{H}\,=\,\Op{H}_g \otimes |G\rangle\langle G|\,~+~
\Op{H}_e \otimes |E\rangle\langle E|
-\Op{\mu}\otimes(|G\rangle\langle E|+|E\rangle\langle G|)\cdot \epsilon(t)\,
\end{equation}
where $|G\rangle$ and $|E\rangle$ are the ground
and excited electronic states and 
$\Op{H}_g$ and $\Op{H}_e$ are the corresponding  vibrational Hamiltonians. 
The electronic surfaces are coupled by the transition dipole operator $\Op{\mu}$, 
controlled by the shaped field $\epsilon(t)$. 

The present model is a simplification of the  Na$_2$ Hilbert space restricting the
number of vibrational levels. On the ground $X^1\Sigma^+_g$ electronic state
the first $40$ vibrational levels selected out from the $66$ bound states are used.
In the excited  $A^1\Sigma^+_u$ state, the lowest $20$ vibrational states are used
out of the $210$ bound levels. The vibrational Hamiltonians become therefore
\begin{equation}\label{eq:hamhad0}
\Op{H}_g=\sum_{i=1}^{40} E_{gi} 
|g_i\rangle \langle g_i|\,;\;\;\;\;\;\;
\Op{H}_e=\sum_{j=1}^{20} E_{ej} |e_j\rangle \langle e_j|\,.
\end{equation}
For Na$_2$ the $00$ transition frequency between the ground vibrational levels of each
surface is $\Omega\equiv E_{e1} - E_{g1}\approx 0.06601 a.u.$ 
($\sim 1.8 eV$ ).
A transition dipole  operator independent of the internuclear distance $R$ 
was considered,
$\Op{\mu}=\mu_0 (|G\rangle\langle E|+|E\rangle\langle G|)$.
This model is
sufficient for the illustrative purpose of demonstrating the execution of 
an algorithm in a molecular setting.

The $N=2^Q$ first levels of the ground electronic surface 
are chosen as the registers representing the $Q$ qubits.
The unitary transformation implemented is a Fourier transform
\cite{WPFLC01} invoked on the $N$ levels on the  $X^1\Sigma^+_g$ electronic state
representing the qubit(s).
The unitary transformation is implemented through
transitions between the two electronic manifolds Cf. Fig. \ref{fig:schme}.

An implementation of the iterative algorithm is chosen where the
$|g_i\rangle\otimes|G\rangle$ and $|e_j\rangle\otimes|E\rangle$
eigenstates are used as the basis $\{|m\rangle\}$.
The $N=2^Q$ first states in the lower surface are used as the 
basis $\{|n\rangle\}$ of the relevant subspace. 
The first $N-1$ energy levels plus the linear combination
$\sum_{n=1}^{N}|n\ra/\sqrt{N}$ are used as the set $|l\rangle$
for the state to state formulation.
The wavefunction  propagations were carried out by
using a Newton polynomial integrator 
\cite{Kosloff94}. The final time for the implementation
is $T=4.5\times10^4 a.u. $ ($\approx$ 1 $psec$). 
In all the cases a Gaussian shape function 
$s(t)=\exp\{-32(t/T-1/2)\}$ and a guess field
$\epsilon_{guess}(t)= \epsilon_0 s(t)\cos(\Omega t)$ 
were chosen.


\begin{figure}[h]
\vspace{1.2cm}
\hspace{0.16\textwidth}
\psfig{figure=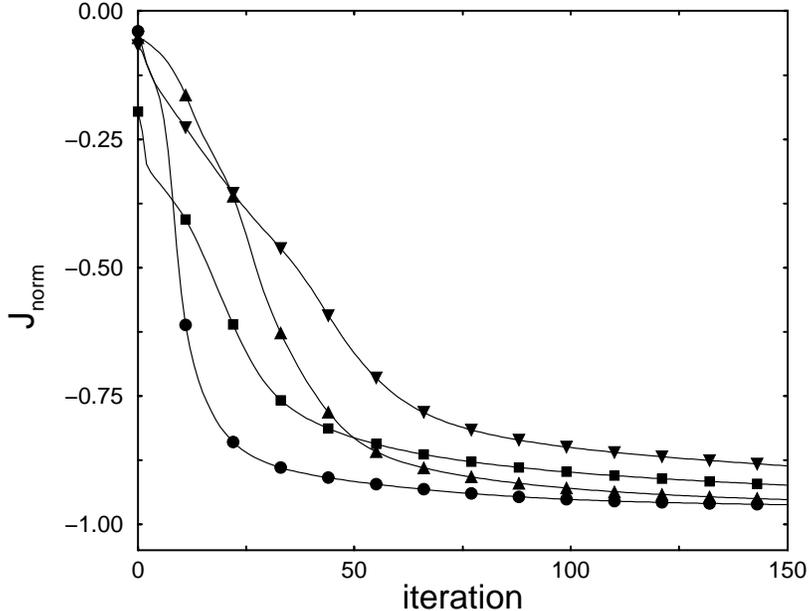,width=0.65\textwidth}
\vspace{.3cm}
\caption{Normalized functional, $J_{norm}$ versus the number of iteration: 
$F_{re}$ (squares), $F_{sm}$ (circles),
$F_{ss}$ (triangles up) for implementing a FFT in $4$ levels. 
The line with triangles pointing down corresponds to 
$F_{ss}$ functional  when $\{|l\rangle\}$ is chosen as
the orthogonal basis $\{|n\rangle\}$. The objective is reached when $J_{norm}=-1.$
$\lambda_0=10^3$ and $\epsilon_0=5\times 10^{-3} a.u.$
in all the cases.
} 
\label{fig:objective}
\end{figure}

\begin{figure}[h]
\vspace{1.2cm}
\hspace{0.16\textwidth}
\psfig{figure=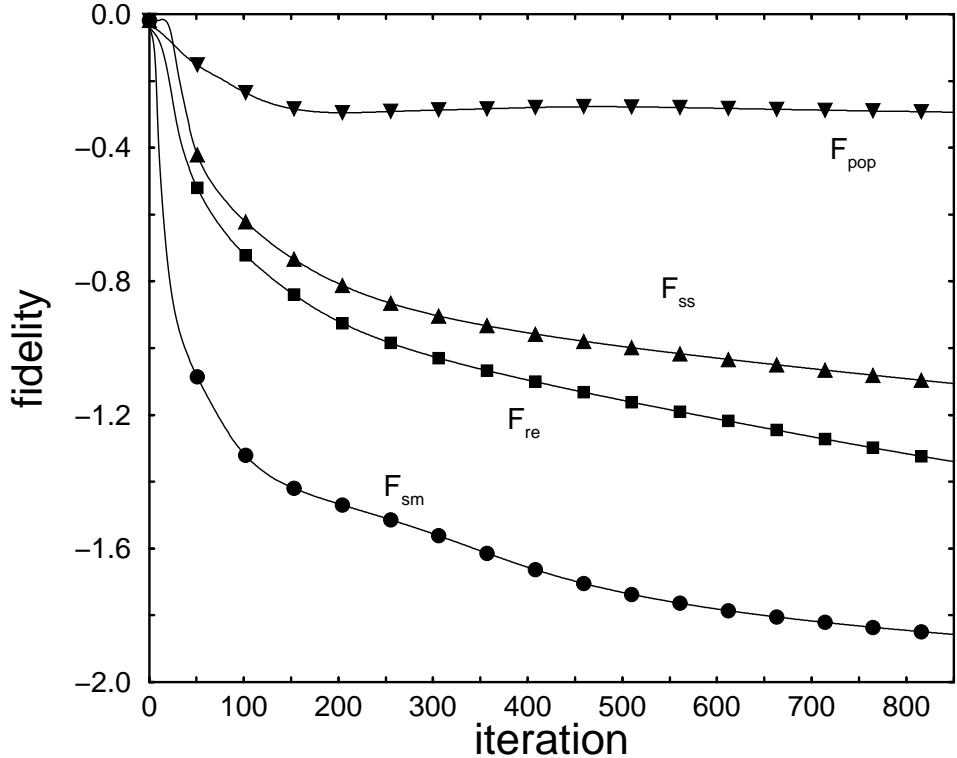,width=0.65\textwidth}
\vspace{.3cm}
\caption{Fidelity of the implementation of the 2 qubit Fourier transform
versus the number of iterations $N_{it}$ 
for the optimization in Fig. \ref{fig:objective}.
The lines correspond to
$F_{re}$ (squares), 
$F_{sm}$ (circles),
and $F_{ss}$ (triangles up). 
$F_{pop}$ (triangles down) denotes the case when
the set $\{|l\rangle\}$ is chosen as
the orthogonal basis $\{|n\rangle\}$ for the functional $F_{ss}$.}
\label{fig:fidelity}
\end{figure}

The implementation of the Fourier
transform in $2$ qubits ($N=4$) embedded in the set of 60 levels is used
for comparing the performance of the methods.
Fig. \ref{fig:objective} shows the change in the normalized
functional, defined as $J_{norm}\equiv J/N$ for $F_{re}$ and $F_{ss}$, 
and $J_{norm}\equiv J/N^2$ for $F_{sm}$, with the progression of the
iterative algorithm. In all the cases the target value
of the normalized functional is $-1$. A large reduction
in the value of the functionals is accomplished in a small number of
iterations.
Notice the behavior of the simultaneous state to state
formulation $F_{ss}$ with an insufficient choice of the
states $|l\rangle$. The algorithm finds a minimum of the
objective, but, as shown in Fig. \ref{fig:fidelity}, the fidelity saturate
at a very low value meaning that 
this field does not generate the target unitary transformation.

\begin{figure}[h]
\vspace{1.2cm}
\hspace{0.16\textwidth}
\psfig{figure=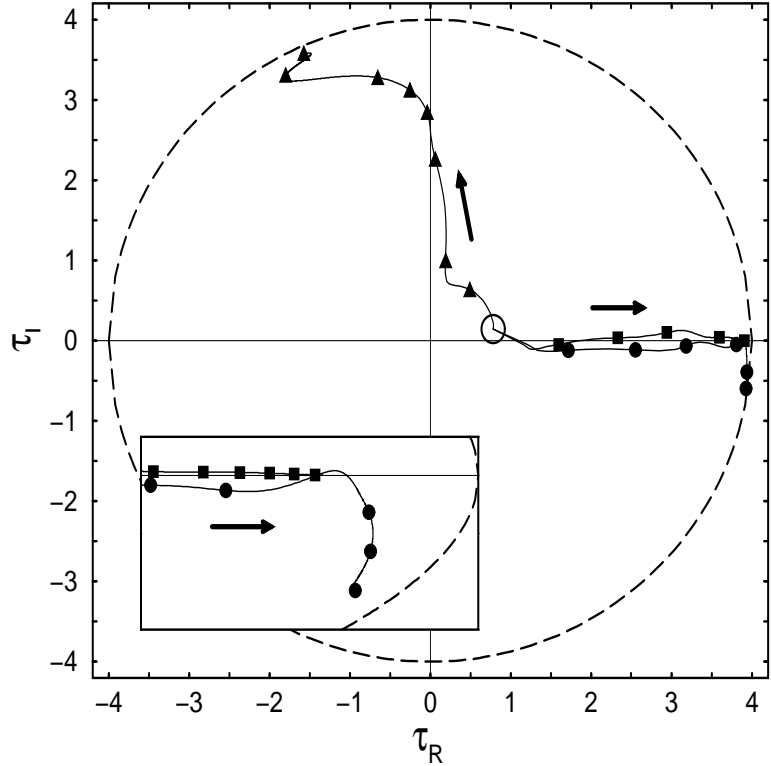,width=0.65\textwidth}
\vspace{.3cm}
\caption{Evolution of the optimization in the 
complex $\tau$ plane for the case in
Fig. \ref{fig:objective}. The lines
correspond to $F_{sm}$ (circles),
$F_{re}$ (squares), 
and $F_{ss}$ (triangles up). The open circle indicates the value of $\tau$
for the common guess field.
The dashed black line is the circle $|\tau|=N$ indicating the target of the methods.
The arrows mark the direction of convergence. The insert enlarges the region
corresponding to the real axes  close
to the circumference.}
\label{fig:tau}
\end{figure}

Fig. \ref{fig:tau} shows the value of $\tau$ for the
field obtained in each iteration. The 
same initial guess was used in all the cases which constituted  
the starting point for all the iterative  optimizations. 
However, the final results depend on the particular functional used. 
As discussed before the method based on $F_{re}$ finds a
solution with a phase factor $\exp{(-i\phi)}\approx+1$.

For the purpose of quantum computing the target
unitary transformation has to achieve
high accuracy. The fidelity functional
\begin{equation}
{\rm fidelity}\,=\,{\rm log}_{10}(1-|\tau|^2/N^2)\,.
\end{equation}
is used to indicate the quality of the solution.
Fig. \ref{fig:fidelity} shows the improvement of the 
fidelity versus the iteration.
The square modulus functional $F_{sm}$ Eq. (\ref{eq:Fsm})
shows a faster convergence rate than the other 
two functionals.

\begin{figure}[h]
\vspace{1.2cm}
\hspace{0.16\textwidth}
\psfig{figure=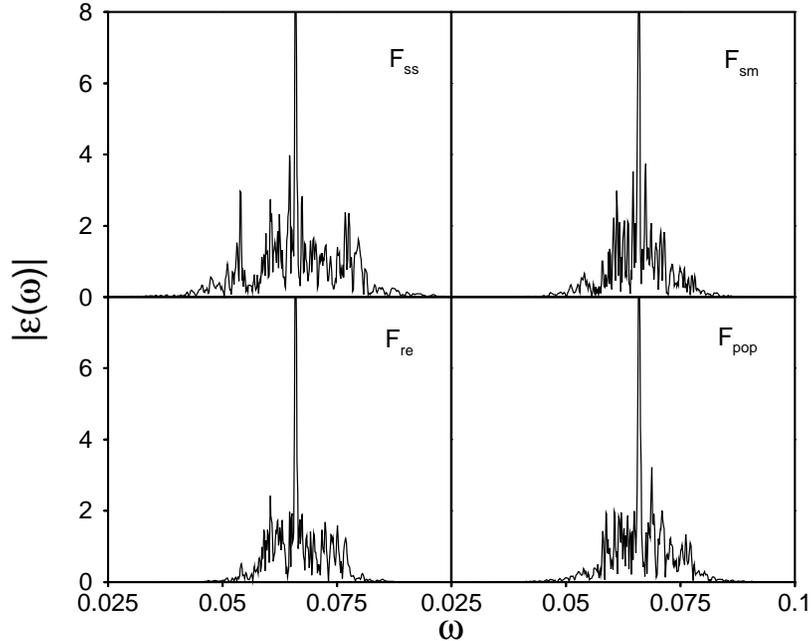,width=0.65\textwidth}
\vspace{.3cm}
\caption{Fourier transform of the optimal field result of the 
optimization in Fig. \ref{fig:fidelity}
for the functionals $F_{re}$,
$F_{sm}$, and $F_{ss}$.} 
\label{fig:frequency}
\end{figure}
 
In Fig. \ref{fig:frequency} the Fourier transform of the field
for each of the optimization procedures is shown.
The large peak at the frequency $\Omega$, seen in all cases,
is the result of the choice of the guess field.
Besides a similar width in frequencies is found. However,
the fidelity reached by the solution corresponding to the 
square modulus functional $F_{sm}$
is significantly better than in the other cases for the same number
of iterations.

\begin{figure}[h]
\vspace{1.2cm}
\hspace{0.16\textwidth}
\psfig{figure=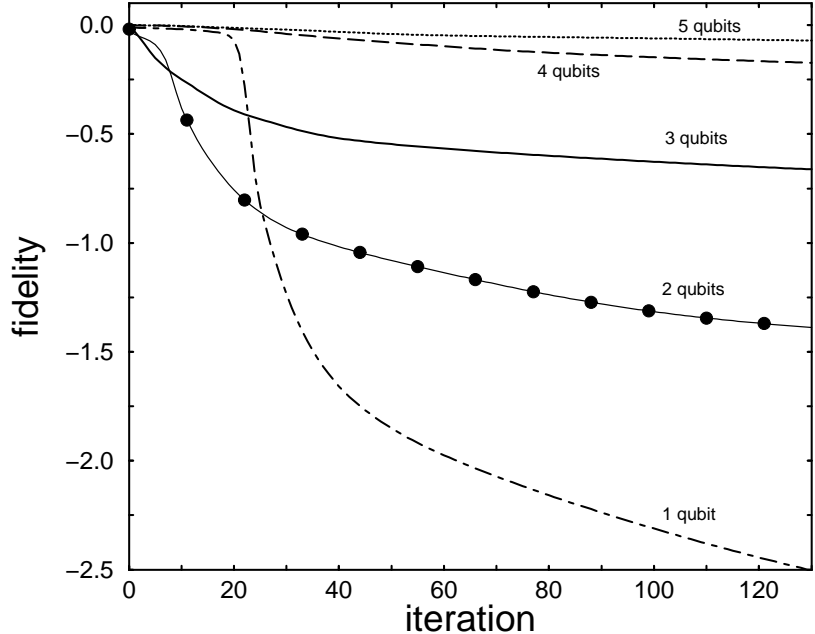,width=0.65\textwidth}
\vspace{.3cm}
\caption{Fidelity versus the number of iterations for implementing a
Fourier transform in $2$ (dashed-dotted line),
$4$ (circles), $8$ (solid line), $16$ (dashed line) and
$32$ (dotted line) levels.
} 
\label{fig:qnumber}
\end{figure}
The molecular model is also used to compare the convergence of
the unitary transformation with the size $N$ of the relevant subspace.
Fig \ref{fig:qnumber} shows the improvement in the fidelity 
versus the number of iterations for implementing a Fourier transform
in  $2$, $4$, $8$, $16$, and $32$ levels ($1$, $2$, $3$, $4$, and $5$
qubits respectively).
The convergence characteristics in the initial iterations strongly depends
on the initial guess and the parameter $\lambda_0$. For example the initial
guess seems inappropriate for the $1$ qubit case which displays an initial
very slow convergence until after 25 iterations it find the right track.
After a large number of iterations  the convergence characteristics settled
meaning that each new iteration was only a slight improvement on the previous one.
As the iteration proceeds the rate of convergence decreases in all cases, 
scaling approximately as the inverse of the number of iterations.
Comparing the rate of convergence for the different number of qubits
after a large number of iterations  the rate seems to be inversely
proportional to the number of levels. High fidelity was
obtained for  $1$, $2$, $3$ qubit cases by continuing to $600$ iterations.
The results allow to compare the integrated intensity of the optimal field:
\begin{equation}
\label{eq:intense}
{\cal I } ~=~ \int_0^T |\mu_0 \epsilon (t)| dt
\end{equation}
The initial integrated intensity for all cases was identical. The optimization
procedure changed ${\cal I}$ depending on the number of qubits.
The converged results show a moderate increase of ${\cal I}$ with the number of levels
starting from  ${\cal I}=42$ for $1$ qubit to ${\cal I}=54$ for $2$ qubits and
${\cal I}=78$ for $3$ qubits.


\section{Discussion}\label{se:conclusions}

An implicit assumption in the optimization procedure is that the system is 
controllable. This means that a field $\epsilon (t)$ exists which implements the 
unitary transformation up to a pre-specified tolerance. The problem
of controllability has been the subject of several studies 
\cite{tarn,Ramakrishna95,Ramakrishna00,TR01,schirmer02}. 
In the context of unitary transformations it has been shown \cite{Ramakrishna95} 
that if the commutators of the operators $\Op H_0$ and $\Op \mu$ 
generate the complete Lie group $SU(N)$, the system
is completely controllable. In more concrete terms addressing the Na$_2$ model, it is 
expected to be completely controllable. 
The reason is that the energy levels are non degenerate and in addition 
each transition is distinct, characterized by a different 
Frank Condon factor $\langle e_j |\Op \mu | g_i \rangle $. 
This controllability property 
will be true in almost any non-symmetric molecular system.

A far reaching conclusion is therefore that for any unitary transformation
contained in the Hilbert space of the molecule, there is a driving
field that implements the transformation in one step. In a molecular system
this task could be achieved in a time scale of a picosecond.
Since a field that executes  such a unitary transformation exists,
how difficult  is it to find  it? Does this optimal field 
have reasonable  intensity and bandwidth?

The OCT scheme can be considered as a classical algorithm employed for the 
inverse problem of finding the field that generates a predefined  unitary transformation. 
The difficulty of the inversion process is related to the
scaling properties of the numerical effort with respect to the number of 
levels $N$.  The best OCT algorithm based on the  $F_{sm}$ functional is 
then used for estimating the scaling.

Simulating the quantum evolution
is the major numerical task of the algorithm implementing OCT.
The basic step is a single vector matrix multiplication which represents
the operation of the Hamiltonian on the wave function. This task scales
as ${\cal O}(M^2)$ for direct vector-matrix multiplication or ${\cal O}(M\log M)$
for grid methods based on FFT \cite{k56}.
The time propagation requires $N_t$ steps which scale as ${\cal O}(T \Delta E)$,
where $\Delta E$ is the energy range of the problem.

The simulation of a unitary transformation in the relevant subspace turns out to be 
$N$ times more costly. Summarizing, the numerical cost of the classical simulation 
of the quantum propagation scales as 
$Cost\sim{\cal O}(2^Q M^2 T \Delta E)$. This scaling relations is consistent
with the fact that a classical
simulation of a quantum unitary transformation scales exponentially with the 
number of qubits.

The numerical cost of the OCT iterative algorithm used for inversion can  now be examined. 
The crucial question is how many iterations are required to obtain
the field that implements the unitary transformation up to a specified fidelity $f$. 
The analysis of the results of Sec. \ref{se:qubits} show that only
the initial iterative steps are very sensitive to the choice of the initial
guess field. Eventually the rate of convergence reaches an asymptotic behavior
where the fidelity becomes inversely proportional
to the number of the iterations steps. In addition Cf. Fig. \ref{fig:qnumber},
the rate of convergence is also inversely proportional to the number of 
levels.
This relation implies that the number
of iterations $N_{it}$ required to achieve the fidelity $f$ becomes 
\begin{equation}
N_{it}\approx b e^{\frac{2^Q |f| }{ a}}\,,
\label{eq:scaling}
\end{equation}
where the coefficients $a$ and $b$ are positives. The data confirm that the
coefficient $a$ is independent of the number of levels $N$. 
The consequence of Eq. (\ref{eq:scaling})
is that the numerical resources required on a classical computer
in order to implement the proposed scheme, scale exponentially with the number of levels $N$.
Finding the field that implements in a single step a large 
unitary transformation is therefore  prohibitively expensive.
Thus fields that achieve high fidelity are only feasible for unitary transformations with
a small relevant subspace. The limiting case would be the one dimensional
state-to-state optimization.
 
Quantum control is based on interferences between many distinct pathways
\cite{RZ00}. State to state coherent control finds a constructive interference
which leads exclusively to the final state. The controllability depends
on having a sufficient amount of interference pathways. 
Implementing a unitary transformation by interferences 
is more complex. In this case the
interference pathways from one state to another have to avoid 
other interference paths which connect other states. 
The possible number of interference pathways becomes
the crucial resource that allows to generate the transformation.

For weak fields, the number of pathways connecting two states in the subspace
is linearly related to the number of auxiliary states on the excited surface.
Practically the bandwidth of the pulse determines this number.
This means that the bandwidth in a weak field implementation of a unitary transformations
has to increase exponentially when the number of levels $N$ increases.
The picture is completely altered when the intensity is allowed to increase.
Rabi cycling increases the number of interference pathways exponentially. The number
of Rabi cycles can be estimated from the integrated intensity $J_{Rabi} \sim {\cal I}/2 \pi$
Cf. Eq. (\ref{eq:intense}), which leads to an estimation
of the number of interference pathways 
${\cal O}(M^{J_{Rabi}})\sim {\cal O}(M^{{\cal I}/2 \pi})$. 
This estimation is consistent with the results of Sec. \ref{se:qubits} where only 
a moderate increase in ${\cal I}$ was observed when the number of qubits 
in the transformation increased. The estimated number of Rabi cycles changed from 
$J_{Rabi} \sim 6$ for $Q=1$ to $J_{Rabi}\sim 8$ for $Q=2$ to $J_{Rabi} \sim 12$ for $Q=3$.
This means that the increase in resources of 
implementing a unitary transformation with Q qubits in a molecular
environment will scale with a low power of  $T\Delta E$ where
$\Delta E$ is the pulse energy. 

In summary,

\begin{itemize}

\item{ A unified approach for obtaining the field that implements a unitary
transformation has enabled the assessment of various formulations.  In addition,
a new algorithm based on the square modulus of $\tau$ was developed. This 
scheme was found to have superior convergence properties with respect
to the number of iterations.}

\item{A unitary transformation could be implemented in a molecular
environment in a time scale of picosecond with reasonable 
bandwidth and intensity. For intense filed conditions the physical resources
scale moderately with the number of qubits in the transformation.}

\item{The inversion problem of finding the field that induces a unitary
transformation seem to be a hard numerical problem 
scaling unfavorably with the number of levels in the transformation.}

\end{itemize}

\section*{Acknowledgments}

J. P. Palao acknowledges financial support of the Gobierno
de Canarias.
This work was supported by Spanish MCT BFM2001-3349,
Gobierno de Canarias PI2002-009 and the Israel Science
Foundation. The Fritz Haber Center is supported by the
Minerva Gesellschaft f\"ur die Forschung, GmbH M\"unchen,
Germany. We thank Christiane Koch for her assistance 
and encouragement. 
Also we thank Zohar Amitay,  David Tannor, Shlomo Sklarz,
and Lajos Diosi for helpful discussions.


\appendix
\section{The variational method}\label{ap:varapp}

An alternative to the Krotov method of optimization is 
the variational method \cite{PDR88,RZ00}.
This method has been used previously in the simultaneous
$N$ state-to-state transitions formulation \cite{TV02}
and for the evolution operator formulation using the 
functional $F_{re}$ \cite{PK02}. In the last case
the variational method was generalized in terms of the 
evolution equation for the unitary transformation.
Unlike the Krotov method the variational
method does not offer a direct algorithm to minimize $F_{sm}$.
For simplicity only the
optimization of the functional $F_{ss}$ is discussed. 
The variational method is based on the functional \cite{TV02}
\begin{eqnarray}\label{app}
&&K(\{\psi_{il},\psi_{fl}\},\Delta\epsilon)=
\sum_{l=1}^{N}\,|\langle\psi_{il}(T)|\,\Op{O}\,|l\rangle|^2
\,-\,\int_{0}^{T} \frac{\lambda_0}{s(t)}\,
|\Delta\epsilon|^2\,dt\nonumber\\
&&-2\,{\rm Re}\left[
\sum_{l=1}^{N}\,\langle\psi_{il}(T)|\,\Op{O}\,|l\rangle
\int_{0}^{T} \langle\psi_{fl}(t)|
\left(\frac{d}{d t}+\frac{i}{\hbar}
\Op{H}(\tilde{\epsilon}+\Delta\epsilon)\right)
|\psi_{il}(t)\rangle\right]\,,
\end{eqnarray}
with the additional condition 
$|\psi_{il}(t=0)\rangle=|l\rangle$. The set of states
$\{| l\rangle\}$ and the target unitary transformation
$\Op{O}$ were introduced in section II. 
$\{|\psi_{il}(t)\}$ denotes the initial states
driven by the field to the final
states $\Op{O}|l\rangle$. The terms $|\varphi_{fl}(t)\rangle$
are interpreted as Lagrange multipliers used as a constraint
to impose  the Schr\"odinger equation.
The two first terms are equivalent to the 
functional (\ref{eq:functional1}) of the Krotov method.
The parameter $\lambda_0$ is now interpreted as a Lagrange multiplier.
The  functional (\ref{app}) differs from the common formulation of OCT
in the form of the field term $\tilde{\epsilon}+\Delta\epsilon$.
$\tilde{\epsilon}$ is a reference field and $\Delta\epsilon$
must be interpreted as the correction used to converge to the optimal
field that implements the target unitary transformation. 
Setting $\tilde{\epsilon}=0$ and interpreting 
$\Delta\epsilon$ as the field the common form is re-attained.

By applying the calculus of variations, requiring $\delta K=0$, with respect to
each element of the set $\{\psi_{il}(t)\}$, 
the evolution equations are reconstructed
\begin{equation}
\frac{d}{dt}|\psi_{il}(t)\rangle\,=\,
-\frac{i}{\hbar}\Op{H}(\tilde{\epsilon}+\Delta\epsilon)
|\psi_{il}(t)\rangle\,,
\end{equation}
with the condition $|\psi_{il}(t=0)\rangle=|l\rangle$
and formal solution 
$|\psi_{il}(t)\rangle=
\Op{U}(t,0;\tilde{\epsilon}+\Delta\epsilon)|l\rangle$.
The variations with respect to the set $\{\psi_{fl}(t)\}$ 
gives 
\begin{equation}
\frac{d}{dt}|\psi_{fl}(t)\rangle\,=\,
-\frac{i}{\hbar}\Op{H}(\tilde{\epsilon}+\Delta\epsilon)
|\psi_{fl}(t)\rangle\,,
\end{equation}
with the condition 
$|\psi_{fl}(t=T)\rangle=\Op{O}|l \rangle$.
The formal solution is 
$|\psi_{fl}(t)\rangle=
\Op{U}(t,T;\tilde{\epsilon}+\Delta\epsilon)\Op{O}|l\rangle$.
Finally, variations with respect to $\Delta\epsilon$ lead to
the correction to the field
\begin{equation}\label{eq:newevar}
\Delta\epsilon(t)\,=\,
-\frac{s(t)}{\lambda_0\,\hbar}
{\rm Im}\left[\sum_{l=1}^{N}\,b_l\,
\langle l |\Op{O}^\dagger\Op{U}^\dagger(t,T;
\tilde{\epsilon}+\Delta\epsilon)\,\Op{\mu}\,
\Op{U}(t,0;\tilde{\epsilon}+\Delta\epsilon)|l\rangle
\right]\,,
\end{equation}
with
\begin{equation}
b_l\,=\,\langle l|\Op{U}^\dagger
(T,0;\tilde{\epsilon}+\Delta\epsilon)\Op{O}|l\rangle\,.
\end{equation}
The correction to the field (\ref{eq:newevar}) is the 
starting point of the iterative algorithms to find the 
optimal field. In such a case the correction to the field
is implicit in the backwards and forwards propagation
of the states in $\Delta\epsilon$.
Several iterative methods have been proposed  \cite{ZBR98}.
In the simplest approach, a guess field $\epsilon^{(0)}$
is used to evaluate $\Delta\epsilon$, that will be used
to obtain the input field in the next iteration. Usually 
it does not converge. 
An alternative procedure \cite{ZBR98} is to evaluate 
$\Op{U}^\dagger$ in Eq. (\ref{eq:newevar}) using the field
in the previous iteration and then to simultaneously obtain
the correction to the field and evaluate $\Op{U}$ with the new field.
This iterative algorithm is identical to the one obtained
from the Krotov method in Sec. \ref{se:optimization}.
A study comparing different iterative algorithms based
on the Krotov and variational methods for the problem of 
state-to-state optimization is described in Ref. \cite{MT03}.


\end{document}